\documentclass[twocolumn]{aastex631}
\shortauthors{McGloughlin, Martins, Steltner et al.}
\shorttitle{E@H O3 bucket search}

\usepackage{ulem}
\usepackage{cellspace}
\usepackage{amssymb}
\usepackage{amsmath}
\usepackage{mathtools}
\usepackage{longtable}
\usepackage{tabu}
\usepackage{color}
\usepackage{etoolbox}
\usepackage{supertabular}
\usepackage{savesym}
\savesymbol{tablenum}
\usepackage{siunitx}
\restoresymbol{SIX}{tablenum}
\usepackage{dcolumn}
\usepackage{wasysym}
\usepackage{comment}

\newcommand{\paramtotaltemplates}{\num{2.2e+19}}
\newcommand{\paramtotalWUsmillions}{\num{10.0}}

\newcommand{\nCandPerWUfiftymHz}{\num{6250}}

\newcommand{\paramWUtotaltemplates}{\num{2.2e+12}}
\newcommand{\paramfmax}{250.0}
\newcommand{\paramfmin}{30.0}

\newcommand{\TcohFUZero}{\num{244}}
\newcommand{\dfreqmuHzFUZero}{\num{1}}
\newcommand{\dfdotfgTenfHzFUZero}{\num{30}}
\newcommand{\mskyFUZero}{\num{7.5e-5}}

\newcommand{\aTwoFrFUOne}{9.8}
\newcommand{\aBSGLtLrFUOne}{-1}

\newcommand{\TcohFUOne}{\num{244}}
\newcommand{\NCandFUOne}{\num{13828991}}
\newcommand{\NSegFUZero}{18}
\newcommand{\NSegFUOne}{18}
\newcommand{\mskyFUOne}{\num{4e-5}}

\newcommand{\dfreqmuHzFUOne}{\num{1}}

\newcommand{\dfdotfgTenfHzFUOne}{5.7}
\newcommand{\meanMismatchFUZero}{48}
\newcommand{\meanMismatchFUOne}{36}

\newcommand{\NSigSOne}{2667}
\newcommand{\SigRecPercent}{\num{89}}

\newcommand{\SRfreqmuHzFUOne}{300}
\newcommand{\SRfreqmuHzFUTwo}{\num{230}}
\newcommand{\SRfdotFUOne}{\num{6000}}
\newcommand{\SRfdotFUTwo}{\num{3800}}
\newcommand{\SRskyFUOne}{22}
\newcommand{\SRskyFUTwo}{21}

\newcommand{\TcohFUTwo}{\num{244}}
\newcommand{\TcohFUThree}{Coh.}

\newcommand{\NSegFUTwo}{18}
\newcommand{\NCandFUTwo}{\num{6556928}}
\newcommand{\NStarFUTwo}{\num{5.2e+09}}
\newcommand{\NStarFUThree}{-}
\newcommand{\NSegFUThree}{1}
\newcommand{\NCandFUThree}{\num{6556928}}
\newcommand{\NCandFUFour}{4}

\newcommand{\mostStringentUL}{\num{6.5e-26}}
\newcommand{\mostStringentULFreq}{\num{173}}
\newcommand{\hwinjreftimestageseven}{1246070525}
\newcommand{\hwinjfiveid}{5}
\newcommand{\hwinjfivealpha}{\ensuremath{20{:}10{:}30.3939}}
\newcommand{\hwinjfivedelta}{\ensuremath{-83{:}50{:}20.9036}}

\newcommand{\hwinjfivefreq}{\num{52.808324358}}
\newcommand{\hwinjfivefonedot}{\num{-4.03e-18}}
\newcommand{\hwinjfivedfreq}{\num{-1.5e-09}}
\newcommand{\hwinjfivedfonedot}{\num{-4.0e-15}}
\newcommand{\hwinjfivedsky}{\ensuremath{0{:}00{:}01.1318}}
\newcommand{\hwinjthreeid}{3}
\newcommand{\hwinjthreealpha}{\ensuremath{11{:}53{:}29.4178}}
\newcommand{\hwinjthreedelta}{\ensuremath{-33{:}26{:}11.7687}}

\newcommand{\hwinjthreefreq}{\num{108.857159393}}
\newcommand{\hwinjthreefonedot}{\num{-1.46e-17}}
\newcommand{\hwinjthreedfreq}{\num{2.9e-09}}
\newcommand{\hwinjthreedfonedot}{\num{-1.4e-14}}
\newcommand{\hwinjthreedsky}{\ensuremath{0{:}00{:}01.7698}}
\newcommand{\hwinjelevenid}{11}
\newcommand{\hwinjelevenalpha}{\ensuremath{19{:}00{:}23.3600}}
\newcommand{\hwinjelevendelta}{\ensuremath{-58{:}16{:}19.5384}}

\newcommand{\hwinjelevenfreq}{\num{31.42469985}}
\newcommand{\hwinjelevenfonedot}{\num{-5.07e-13}}
\newcommand{\hwinjelevendfreq}{\num{2.4e-08}}
\newcommand{\hwinjelevendfonedot}{\num{4.2e-15}}
\newcommand{\hwinjelevendsky}{\ensuremath{0{:}00{:}03.7466}}

\newcommand{\ie}{i.\,e.\@ }

\newcommand*{\defeq}{\mathrel{\vcenter{\baselineskip0.5ex \lineskiplimit0pt
                     \hbox{\scriptsize.}\hbox{\scriptsize.}}}%
                     =}

\newcommand*\diff{\mathop{}\!\mathrm{d}}

\newcommand{\dparams}{\ensuremath{\vartheta}}

\newcommand{\fstat}{\ensuremath{\mathcal{F}}}

\NewDocumentCommand{\lratio}{ e{^} s o >{\SplitArgument{1}{|}}m }{%
    \operatorname{\Lambda}
    \IfValueT{#1}{{\!}^{#1}}
    \IfBooleanTF{#2}{
        \expectarg*{\expectvar#4}%
    }{
        \IfNoValueTF{#3}{
            \expectarg{\expectvar#4}%
        }{
            \expectarg[#3]{\expectvar#4}%
        }%
    }%
}
\NewDocumentCommand{\lik}{ e{^} s o >{\SplitArgument{1}{|}}m }{%
    \operatorname{\mathcal{L}}
    \IfValueT{#1}{{\!}^{#1}}
    \IfBooleanTF{#2}{
        \expectarg*{\expectvar#4}%
    }{
        \IfNoValueTF{#3}{
            \expectarg{\expectvar#4}%
        }{
            \expectarg[#3]{\expectvar#4}%
        }%
    }%
}

\NewDocumentCommand{\probdensity}{ e{_} e{^} s o >{\SplitArgument{1}{|}}m }{%
    \IfValueTF{#1}{f_{#1}}{p}
    \IfValueT{#2}{^{#2}}
    \IfBooleanTF{#3}{
        \expectarg*{\expectvar#5}%
    }{
        \IfNoValueTF{#4}{
            \expectarg{\expectvar#5}%
        }{
            \expectarg[#4]{\expectvar#5}%
        }%
    }%
}

\NewDocumentCommand{\prob}{ e{^} s o >{\SplitArgument{1}{|}}m }{%
    \operatorname{P}
    \IfValueT{#1}{{\!}^{#1}}
    \IfBooleanTF{#2}{
        \expectarg*{\expectvar#4}%
    }{
        \IfNoValueTF{#3}{
            \expectarg{\expectvar#4}%
        }{
            \expectarg[#3]{\expectvar#4}%
        }%
    }%
}

\NewDocumentCommand{\gaussian}{ e{^} s o >{\SplitArgument{1}{|}}m }{%
    \mathcal{G}
    \IfValueT{#1}{{\!}^{#1}}
    \IfBooleanTF{#2}{
        \expectarg*{\expectvar#4}%
    }{
        \IfNoValueTF{#3}{
            \expectarg{\expectvar#4}%
        }{
            \expectarg[#3]{\expectvar#4}%
        }%
    }%
}

\NewDocumentCommand{\unif}{ e{^} s o >{\SplitArgument{1}{|}}m }{%
    \mathcal{U}
    \IfValueT{#1}{{\!}^{#1}}
    \IfBooleanTF{#2}{
        \expectarg*{\expectvar#4}%
    }{
        \IfNoValueTF{#3}{
            \expectarg{\expectvar#4}%
        }{
            \expectarg[#3]{\expectvar#4}%
        }%
    }%
}

\NewDocumentCommand{\cdf}{ e{^} s o >{\SplitArgument{1}{|}}m }{%
    \operatorname{CDF}
    \IfValueT{#1}{{\!}^{#1}}
    \IfBooleanTF{#2}{
        \expectarg*{\expectvar#4}%
    }{
        \IfNoValueTF{#3}{
            \expectarg{\expectvar#4}%
        }{
            \expectarg[#3]{\expectvar#4}%
        }%
    }%
}
\NewDocumentCommand{\fstatistic}{ e{_} e{^} s o >{\SplitArgument{1}{|}}m }{%
     \operatorname{\mathcal{F}}
    \IfValueT{#1}{{\!}_{#1}}
    \IfValueT{#2}{{\!}^{#2}}
    \IfBooleanTF{#3}{
        \expectarg*{\expectvar#5}%
    }{
        \IfNoValueTF{#4}{
            \expectarg{\expectvar#5}%
        }{
            \expectarg[#4]{\expectvar#5w}%
        }%
    }%
}
\NewDocumentCommand{\semcohfstatistic}{ e{^} s o >{\SplitArgument{1}{|}}m }{%
    N_\mathrm{seg,a}\bar{\mathcal{F}}
    \IfValueT{#1}{{}^{#1}}
    \IfBooleanTF{#2}{
        \expectarg*{\expectvar#4}%
    }{
        \IfNoValueTF{#3}{
            \expectarg{\expectvar#4}%
        }{
            \expectarg[#3]{\expectvar#4}%
        }%
    }%
}
\NewDocumentCommand{\expectvar}{mm}{%
    #1\IfValueT{#2}{\nonscript\;\delimsize\vert\nonscript\;#2}%
}
\DeclarePairedDelimiterX{\expectarg}[1]{(}{)}{#1}

\newcommand{\fiftyMHzband}{\SI{50}{\milli\hertz}}

\begin{document}

\title{Einstein@Home all-sky ``bucket" search for continuous gravitational waves in LIGO O3 public data}

\correspondingauthor{B. McGloughlin}
\email{brian.mcgloughlin@aei.mpg.de}
\correspondingauthor{M.A. Papa}
\email{maria.alessandra.papa@aei.mpg.de}

\author[0009-0002-4068-7911]{B. McGloughlin}
\author[0009-0002-3912-189X]{J. Martins}
\author[0000-0003-1833-5493]{B. Steltner}
\author[0000-0002-1007-5298]{M. A. Papa}
\author[0000-0001-5296-7035]{H.-B. Eggenstein}
\author{B. Machenschalk}
\author[0000-0002-3789-6424]{R. Prix}
\author{M. Bensch}
\affiliation{Max Planck Institute for Gravitational Physics (Albert Einstein Institute), D-30167 Hannover, Germany}
\affiliation{Leibniz Universit\"at Hannover, D-30167 Hannover, Germany}

\begin{abstract}
	We conduct an all-sky search for continuous gravitational waves using LIGO O3 public data from the Hanford and Livingston detectors. We search for nearly-monochromatic signals with frequencies $\SI{\paramfmin}{\hertz} \leq f \leq \SI{\paramfmax}{\hertz}$ and spin-down $-2.7 \times 10^{-9}\, \text{Hz/s} \leq \dot{f} \leq 0.2 \times 10^{-9}$ Hz/s. We deploy this search on the Einstein@Home volunteer-computing project and on three super computer clusters;  the Atlas supercomputer at the Max Planck Institute for Gravitational Physics, and the two high performance computing systems Raven and Viper at the Max Planck Computing and Data Facility. Our results are consistent with a non-detection. We set upper limits on the gravitational wave amplitude $h_{0}$, and translate these to upper limits on the neutron star ellipticity and on the r-mode amplitude. The most stringent upper limits are at \mostStringentULFreq{} Hz with $h_{0} = \mostStringentUL$, at the 90\% confidence level.
\end{abstract}

\section{Introduction}
Continuous gravitational waves are expected to be produced by rotating neutron stars due to non-axial symmetry and from the destabilization of r-modes \citep{Owen:1998xg}. Signals may also potentially come from more exotic sources such as inspiraling dark matter or axion-like particles surrounding back holes \citep{Zhu:2020tht}.
The expected continuous gravitational wave amplitude at Earth is several orders of magnitude smaller than that of commonly detected signals from compact binary coalescence events. Since the signal is long-lasting ($\sim$ years) one can integrate the detector data over many months and increase the signal-to-noise ratio (SNR) significantly.

The number of resolvable waveforms grows very quickly with the observation time $T{_\textrm{obs}}$. As a result, the most computationally demanding continuous gravitational wave searches are those which aim to detect signals from unknown sources, as the parameter space to investigate is large. Broad-frequency-band all-sky searches fall in this category and the search set-up requires decisions on how to balance search \emph{depth} and search \emph{breadth}.

In this paper we concentrate on a relatively small frequency band -- 220 Hz -- where the LIGO detectors are the most sensitive \citep{aLIGO:2020wna}, and there aim at the highest possible sensitivity. We achieve a $25\%$ sensitivity improvement with respect to the previous low- and high-frequency Einstein@Home all-sky searches on public O3 data \citep{Steltner:2023cfk,McGloughlin:2025eso}, which are the most sensitive searches to date probing a spin-down range at least as wide.

Approximately $14$ million candidates from the initial search are followed-up. We find our results to be consistent with a non-detection and we set upper limits on the gravitational wave amplitude accordingly.

This paper is organized as follows: Section \ref{sec:Signal} describes the signal model and Section \ref{sec:Generalities} outlines the generalities of this search. Section \ref{sec:EAHSearch} describes the initial stage of the search, while Section \ref{sec:FollowUp Searches} describes the follow-up procedure. In Section \ref{sec:Results}, the results are presented, and we conclude in Section \ref{sec:Conclusions}.

\section{The signal}\label{sec:Signal}

The search described in this paper targets quasi-monochromatic gravitational wave signals described in \citep{Jaranowski:1998qm} that are subject to both frequency and amplitude modulations due to the motion of the Earth. The detector outputs a signal of the form:
\begin{equation}
	h(t) = F_{+}(\alpha, \delta, \psi;t)h_{+}(t) + F_{\times}(\alpha, \delta, \psi;t)h_{\times}(t).
\end{equation}
Where $F_{+}(\alpha, \delta, \psi;t)$ and $F_{\times}(\alpha, \delta, \psi;t)$ are called the detector beam-pattern functions for both the ``$+$" and ``$\times$" wave polarizations that depend on the inclination and declination of the source $(\alpha,\delta)$, the polarization angle $\psi$ as well as the time at the detector t.
The plus and cross waveforms take the form:
\begin{eqnarray}
	h_{+}(t) = A_{+}\cos\Phi(t) \nonumber\\
	h_{\times}(t) = A_{\times}\sin\Phi(t)
\end{eqnarray}
With amplitudes:
\begin{eqnarray}
	A_{+} &=& \frac{1}{2}h_{0}(1 + \cos^{2}{\iota}) \nonumber\\
	A_{\times} &=& h_{0}\cos{\iota}
\end{eqnarray}
Here, $h_{0} \ge 0$ is the intrinsic gravitational wave amplitude, $0 \le \iota \le \pi$ is the angle between the angular momentum vector of the neutron star and the line of sight and $\Phi (t)$ is the phase of the gravitational wave at time t.

At rest with respect to the source -- for instance at the solar system barycenter (SSB), for an isolated source -- the phase is
\begin{equation}
	\Phi(\tau_{SSB}) = \Phi_{0} + \newline\\ 2\pi[f(\tau_{SSB} - \tau_{0SSB})] + \frac{1}{2}\dot{f}(\tau_{SSB} - \tau_{0SSB})^{2}]
\end{equation}
We take $\tau_{0SSB} = 1246070525.0$ as a reference time.
In the detector frame, the phase-evolution then depends on a set of four parameters, $\dparams = (f, \dot f, \alpha, \delta)$.

\section{Search Generalities}\label{sec:Generalities}
\subsection{The Data}
This search uses Advanced LIGO data from the first half of the third observation run (O3a), i.e. between GPS time 1238166018 (Apr 01 15:00:00 GMT 2019) and 1254150018 (Oct 03 15:00:00 GMT 2019) from the LIGO Hanford (LHO) and Livingston (LLO) Observatory. Data from other gravitational wave detectors is not used due to their lower sensitivity.

Following \citet{Steltner:2023cfk}, as with previous Einstein@Home searches, we remove noise that would degrade the quality of the search; so-called lines and glitches in the frequency and time-domain respectively \citep{Steltner:2021qjy}.

The input to our searches is in the form of short time-baseline Fourier Transforms (SFTs) each of 1 hour duration. For searching less than $\approx 10^4$ frequencies, input SFT from longer time-baselines are more efficient in processing, and the lower maximal frequency of this search allows us to use this longer time-baselines compared to our ``usual'' SFTs of half-hour duration. All data is then split into data segments of duration $T_{\textrm{coh}}$, see Fig.~\ref{fig:NsegSegmentation}.

\subsection{Search Grids}

For a rotating isolated neutron star, the template waveform is defined by four parameters; the frequency, the first frequency derivative (spin-down) and the two sky coordinates of the source. The grids in frequency and spin-down are each described by a single parameter, the grid spacing, which is constant over the search range. The sky grid is approximately uniform on the celestial sphere orthogonally projected on the ecliptic plane. The tiling is an hexagonal covering of the unit circle with hexagon edge length:

\begin{equation}\label{eq:skyGridSpacing}
	d(m_{\textrm{sky}}) = 0.15 \sqrt{m_{\textrm{sky}}}\left[\frac{100 \textrm{Hz}}{f}\right],
\end{equation}
where $m_{\textrm{sky}}$ is a scaling factor which determines the overall coarseness of the sky grid. The product of all search grids is called the template bank, consisting of all template waveforms which will be searched over.

\subsubsection{Search Setup}

The template bank is set up in such a way that for a given computational capacity the average fractional loss in signal-to-noise, the ``mismatch'' $\left<\mu\right>$, due to a signal falling in-between grid-points is minimal.

As in previous searches \citep{Steltner:2023cfk,McGloughlin:2025eso}, the template bank is chosen among a vast number of setups as the most efficient search, i.e. the one that yields the lowest detectable signal at fixed computational cost. We refer to \citet{McGloughlin:2025eso} for a description of the setup optimization scheme.

\subsection{Detection Statistics}

\begin{figure}[h!tbp]
	\includegraphics[width=\columnwidth]{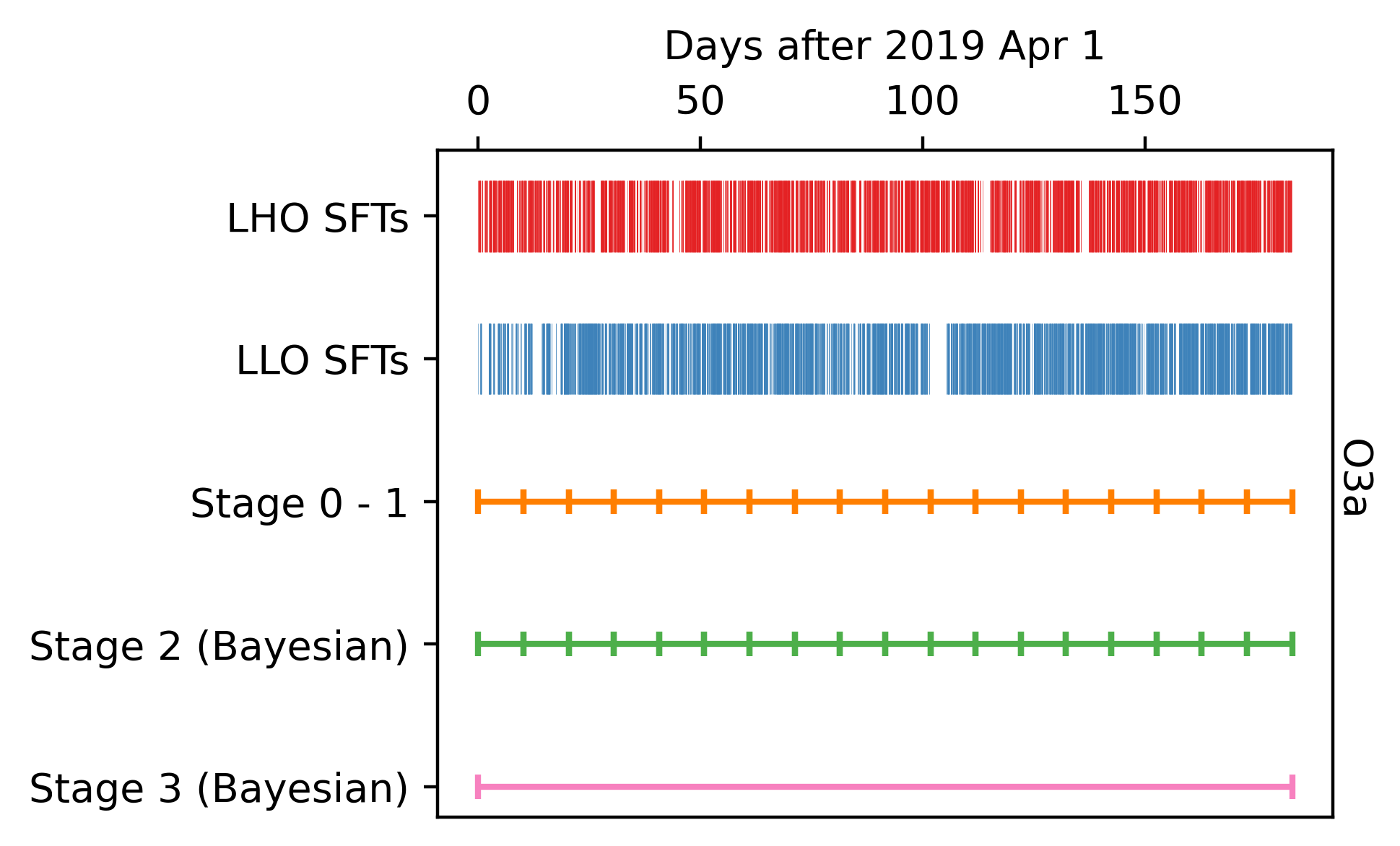}
	\caption{Times of input data (SFTs in LIGO Hanford (LHO) and Livingston (LLO)) and segmentation of that data in the coherent searches of Stage 0 and 1, and the segmentation of the Bayesian search in Stage 2. The Bayesian Stage 3 is a fully-coherent search on O3a data.}
	\label{fig:NsegSegmentation}
\end{figure}

For each search stage the data is partitioned into $N_{\textrm{seg}}$ segments of equal duration $T_{\textrm{coh}}$, see Fig.~\ref{fig:NsegSegmentation}. The data from both detectors in each segment is combined coherently to calculate the detection statistic, the $\mathcal{F}$-statistic \citep{Jaranowski:1998qm,Cutler:2005hc}. The resulting $\mathcal{F}$-statistic from each segment, $\mathcal{F}_{i}$, are summed and averaged to produce the semi-coherent statistic:

\begin{equation}
	\bar{\mathcal{F}} := \frac{1}{N_\mathrm{seg}}\sum_{i=1}^{N_{\mathrm{seg}}}\mathcal{F}_{i}
\end{equation}
For efficiency reasons, the $\mathcal{F}$-statistic is first calculated on a coarse grid in parameter space and then approximated on a finer grid in spin-down \citep{Pletsch:2010xb}. Results with the highest detection statistic values are then recomputed exactly at the fine-grid point.

The $\mathcal{F}$-statistic test the signal hypothesis against a pure Gaussian noise hypothesis and so it is  susceptible to non-Gaussian disturbances. For this reason we also employ the line and transient-line robust detection statistic $\hat{\beta}_{S/GLtL}$ \citep{Keitel:2015ova}, which tests the signal hypothesis against an expanded noise hypothesis, i.e. “G” Gaussian noise or “L” lines or “tL” transient lines.

\section{Einstein@Home Search}\label{sec:EAHSearch}
\subsection{Computational load distribution}

The computational cost of the initial search is split as follows: 30\% is run on the Atlas computer cluster at the Max Planck Institute for Gravitational Physics,  25\% on the Raven and Viper high-performance computing (HPC) systems at the Max Planck Computing and Data Facility and the remaining 45\% on the Einstein@Home project. Einstein@Home\footnote{\url{https://einsteinathome.org}} is a distributed computing project where volunteers can spend their idle computational cycles for scientific computational projects \citep{Boinc2, Boinc3}.

All of these search components are run on Graphical Processing Units (GPUs) with the exception of Viper which uses Accelerated Processing Units (APUs). The \texttt{NVIDIA Geforce RTX 2070 SUPER} and \texttt{NVIDIA RTX 4000 Ada Generation} models are exclusively used on Atlas, the \texttt{NVIDIA A100} model on Raven, the \texttt{AMD Instinct MI300A Accelerators} model on Viper, and a diverse range of GPU models is used on Einstein@Home.

The total number of templates searched on all systems is \paramtotaltemplates. The search is divided into work-units (WUs), where each WU searches approximately 225 sky points, 1 Hz in frequency and the entire spin-down range, totaling \paramWUtotaltemplates{} templates each. A total of \paramtotalWUsmillions{} million WUs are necessary to cover the entire parameter range. The average runtime for one WU on a \texttt{NVIDIA Geforce RTX 2070 SUPER} is $\approx 40$ minutes. Each WU returns a fixed amount of the top-ranking results. To mitigate saturation effects due to e.g. disturbances, each WU returns the \nCandPerWUfiftymHz{} top-ranking results per \fiftyMHzband.

\begin{figure}[h!tbp]
	\centering
	\includegraphics[width=\columnwidth]{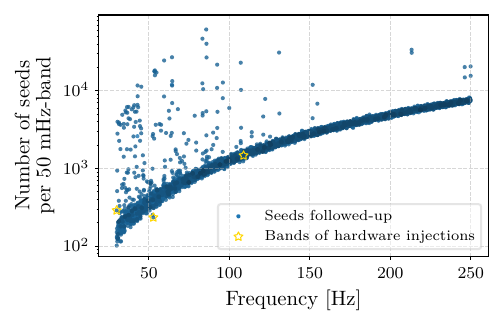}
	\caption{The number of Stage 0 clusters from every \SI{50}{\milli\hertz} band. The relatively large amplitude of the hardware injection in the \SI{50.80}{\hertz} band gives rise to a saturation effect in which most high ranking detection statistic templates are associated with it. This results in fewer, larger clusters being formed in this \SI{50}{\milli\hertz} band overall.}
	\label{fig:seedsPerBand}
\end{figure}

\subsection{Post-processing}\label{subsec:Stage0Post}

The following post-processing steps are performed on the results returned from the Stage 0 search:

\begin{description}
	\item[Banding]
	      all results from different WUs covering the same 50 mHz band are merged together.
	\item[Clustering]
   With the clustering method of \cite{Steltner:2022aze} results due to the same underlying cause are grouped together and considered as a single candidate. Each candidate has the template parameter values of the highest detection statistic result from the cluster, which we refer to as the seed. If the cluster is due to a signal from our target signal population (see next section), with high confidence (over 99\%) the signal parameters ${\dparams}_{s}$ are within a distance $\Delta{\dparams}$ of the seed parameters ${\dparams}_{\textrm{seed}}$. Fig.~\ref{fig:seedsPerBand} shows the number of seeds in each 50 mHz band which grows with frequency as the number of search templates increases with frequency. Clustering produces $\approx 14$ million candidates, which we follow-up. 
	\item[Visual Inspection]
	      All banded results are visually inspected for bands that are deemed likely to be disturbed \citep{LIGOScientific:2017wva}. Bands that are considered disturbed from this examination are excluded from the upper limit procedure but seeds produced from these bands will be followed up.
\end{description}

\section{Follow-Up}\label{sec:FollowUp Searches}
We perform a hierarchical follow-up of the $\NCandFUOne$ candidates identified by our clustering procedure.
Each stage of the search increases in sensitivity by increasing the coherence time $T_\mathrm{coh}$ and/or decreasing the average mismatch $\left<\mu\right>$. Generally, the signal-to-noise ratio of candidates due to a signal increases, while that of candidates due to noise do not. Noise and signal distributions separate, enabling the rejection of false alarms.

We use a reference population of simulated test signals to set up the search hierarchy.
The reference population occupies three half-Hz bands around the frequencies $\qtylist{50;105;225}{\hertz}$.
Within each band, the frequency is drawn from a uniform distribution.
The intrinsic amplitude $h_0$ of all test signals is chosen based on the sensitivity depth \citep{Behnke:2014tma} $\mathcal{D}\approx 70~[{1/\sqrt{\textrm{Hz}}}]$  targeted by this search:
\begin{equation}
	h_{0}(f) = \frac{\sqrt{S^{-1}(f)}}{\mathcal{D}} =\frac{\sqrt{S^{-1}(f)}}{70~[{1/\sqrt{\textrm{Hz}}}]}.
\end{equation}
For the other parameters, we choose log-uniform priors on $\dot f$, isotropic priors on the sky position, and uniform priors on $\cos\iota$, $\psi$ and $\phi_0$.
We apply Stage 0 and clustering to this reference population and recover $\NSigSOne$ ($\SigRecPercent$\%). All subsequent search stages are configured such that we detect all test signals.

Our follow-up is based on two methods: Stage 1 is a template-bank-based refinement stage, while Stages 2 and 3 employ a Bayesian follow-up method.
The setup and results of each stage are given in Table~\ref{tab:FUtable}.

\subsection{The first follow-up stage}
Stage 1 of the follow-up is a refinement stage, which mostly serves to decrease the uncertainty range around the candidates. The same $T_{\rm coh} = 244 \rm h$ as Stage 0 is used, but the smaller parameter space volume of Stage 1 allows for finer resolutions, reducing the mismatch and increasing the sensitivity. Stage 1 was performed using both CPUs and GPUs of the Atlas, Raven, and Viper computing clusters of the Max-Planck-Society.

\begin{figure}
	\includegraphics[width=\columnwidth]{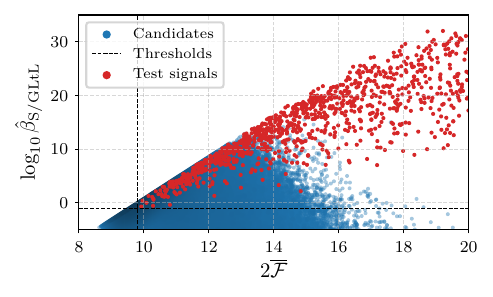}
	\caption{The distribution of detection statistic values of the candidates and the reference signal population at Stage 1. Dashed lines mark the thresholds used to veto candidates. }
	\label{fig:gridFollowUp}
\end{figure}

The most significant result from each Stage 1 search is selected based on the $2\fstat$-statistic. A joint threshold on both $2\fstat$ and $\hat{\beta}_{S/GLtL}$ is used to determine whether the candidate should be further investigated. These threshold are set based on the target signal population, as shown in Fig.~\ref{fig:gridFollowUp}:

\begin{equation}
\begin{cases}
	2\bar{\mathcal{F}}^{\mathrm{thr}} &= \aTwoFrFUOne   \\ 
	\log_{10}\hat{\beta}_{\mathrm{S}/\mathrm{GLtL}}^{\mathrm{thr}} &=  \aBSGLtLrFUOne
\end{cases}
\end{equation}
The reference population is not yet significantly separated from the candidates, and further follow-up is necessary but these thresholds veto $\approx 50\%$ of the candidates.

We estimate a new uncertainty range from the test signal seeds in the same way as described for clustering (Section~\ref{subsec:Stage0Post}).
The estimated uncertainty range is cited in Table \ref{tab:FUtable} as the Stage 2 search range.
Compared to the Stage 1 search regions, the uncertainty ranges are $\approx 4$ times smaller.

The rejection of $\approx 50\%$ of the candidates and the reduction of the uncertainty ranges by $75\%$ leads to an approximate $8 \times$ reduction of the total parameter space volume to search in the next stage.
We note that past searches with similar methodologies have achieved orders of magnitude larger reductions of the parameter space volume to follow-up \citep{Steltner:2020hfd,Steltner:2023cfk,McGloughlin:2025eso}, enabling orders of magnitude finer grids at the next stage at fixed computing cost.
The reason for this is the relatively small decrease in mismatch from Stage 0 to Stage 1, at the same $T_\textrm{coh}$. The consequence is that the follow-up of the Stage 1 candidates of this search remains a significant computational challenge. 

\subsection{The Bayesian follow-up method}

Stages 2 and 3 use a Bayesian follow-up method and were run exclusively on CPUs at the Atlas computing cluster.

The follow-up method is described in detail in \citet{Martins:2025jnq} and also applied in \citet{McGloughlin:2025eso}.
Here, we only give a brief summary.

Given a candidate with uncertainty region ${\dparams}_{\textrm{seed}}\pm \Delta{\dparams}$ to follow-up in Stage $a$, we consider the posterior probability distributions of the phase-evolution parameters $\dparams$ using Bayes' theorem \citep{Prix:2009tq,Martins:2025jnq}:
\begin{equation}
	\label{eq:thetaPosteriorsemcoh}
	\probdensity{\dparams | N_{\mathrm{seg},a}} = \frac{ e^{\semcohfstatistic{\dparams}}\probdensity^a{\dparams}}{\int_{{\dparams}_{\textrm{seed}}\pm \Delta{\dparams}} e^{\semcohfstatistic{\dparams}}\probdensity^a{\dparams}\diff\dparams}.
\end{equation}
Here, $\probdensity^a{\dparams}$ is the prior probability distribution at Stage $a$ and it is determined by the results of the previous stage.
We use the nested sampling algorithm \citep{Skilling:2004pqw}, specifically \texttt{dynesty} \citep{Speagle:2019ivv} wrapped by \texttt{Bilby} \citep{Ashton:2018jfp} to produce samples from the posterior distribution and to estimate the evidence-integral
\begin{equation}\label{eq:evidence}
	Z^a = \int_{{\dparams}_{\textrm{seed}}\pm \Delta{\dparams}} e^{\semcohfstatistic{\dparams}}\probdensity^a{\dparams}\diff\dparams.
\end{equation}
We note that, in practice, we substitute $\semcohfstatistic{\dparams}$ with the estimated signal power 
\begin{equation}
 \hat\rho^2 \defeq 2 \semcohfstatistic{\dparams} - 4 N_{\mathrm{seg}, a}.
\end{equation}
For each candidate and stage, we model the posterior density function with a Gaussian mixture model.

At Stage $2$, we use uninformative priors following Jeffreys' invariance principle. The principle yields uniform priors on $f$ and $\dot f$  and
priors on the sky-position parameters $(\alpha, \delta)$ that are uniform when orthogonally projected onto the ecliptic plane.

The prior at Stage $3$ is the inferred posterior distribution $\probdensity{\dparams | N_{\mathrm{seg},2}}$ from the previous stage:
\begin{equation}
	\probdensity^{3}{\dparams} = \probdensity{\dparams | N_{\mathrm{seg},2}}.
\end{equation}

We use the estimated evidences $Z^3$ as a test-statistic to compare the candidates' results and the test signal population.
We further define a test-statistic based on the (relative) change in evidence compared to the initial Bayesian stage (Stage 2):
\begin{equation}\label{eq:r}
	R^3 = \frac{\log Z^3 - \log Z^2}{\log Z^2}.
\end{equation}
This down-weights in a fully-coherent search noise-features in the data that appear signal-like in a semi-coherent search.

At Stage 2, the width of the uncertainty ranges, the number of seeds selected and the smaller signal amplitudes targeted pose a significant computational challenge for a stochastic follow-up.
This issue is resolved with two additions with respect to \cite{Martins:2025jnq,McGloughlin:2025eso}:
\begin{enumerate}

	\item By default, nested sampling is initialized with $n_\mathrm{live}$ random samples from the prior distribution. Stage 1 did not produce a posterior to be used as the prior here, but the search results contain information. They cannot directly be used as a prior because they are too coarse. We do however usefully employ them to initialize the live points: 
we take the $n_\mathrm{live}$ highest ranking templates from Stage 1 as the initial samples.

	\item 
The $\fstat$-statistic response to continuous wave signals can be highly multimodal, and occasionally all $n_\mathrm{live}$ points lie on secondary modes. We employ differential-evolution slice sampling, $\mathrm{DE}_\mathrm{SL}$ \cite{Buchner:2022}, to improve the chance of nested sampling unveiling further modes, when proposing new live points. This is an enhancement to \texttt{dynesty} and \texttt{Bilby}.
      
\end{enumerate}

With this configuration, all $\NSigSOne$ signals from the reference population are recovered.

\subsection{Bayesian follow-up results}

We apply Stage 2 and 3 to the $\NCandFUTwo$ candidates that survive Stage 1.

Between Stage 2 and 3, no vetoing is applied, and we obtain fully coherent search results for all Stage 2 candidates.

\begin{figure}
	\includegraphics[width=\columnwidth]{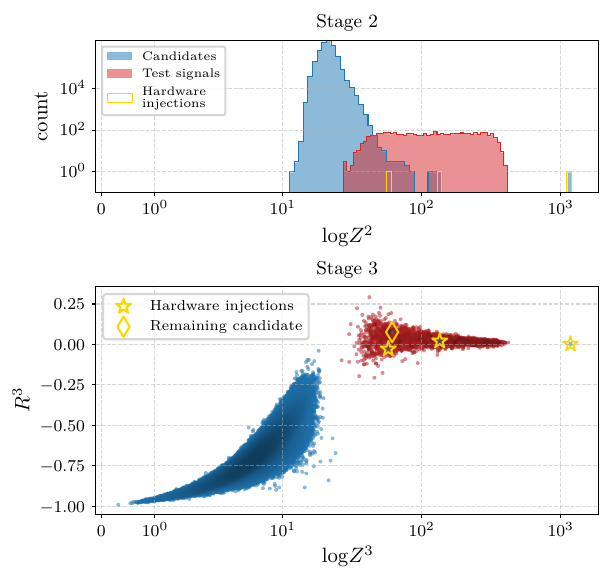}
	\caption{Results of the Bayesian follow-up in Stage 2 and 3. Upper plot (Stage 2): Histograms of the evidences $Z^2$ (Eq.~\ref{eq:evidence}) of the candidates and the test signals.
		Lower plot (Stage 3): Scatter plot of the evidences $Z^3$ and $R^3$ (Eq.~\ref{eq:r}) for candidates and the reference population.}
	\label{fig:bayesianresults}
\end{figure}

The results of the Bayesian follow-up are shown in Fig.~\ref{fig:bayesianresults}.
At Stage 2, the evidences found for the reference population are not yet significant compared to the candidates' results.
At Stage 3, the candidate distribution has clearly separated from the reference population.
However, two noteworthy features of the candidate distribution remain.

First, four candidates have $(Z^3, R^3)$-values that remain compatible with the reference population.
Three of these candidates are due to so-called hardware injections.
The hardware injections and the remaining fourth candidate are described in more detail in Sections~\ref{sec:hardware_injections} and~\ref{sec:remaining_candidate}, respectively.

Second, we note outliers from the general shape of the candidate distribution with higher evidences, but small $R^3$-values.  These candidates correspond to signal-like transients in the data that match our signal model at lower coherences much better than at full coherence.
We do not consider any of these candidates as potential astrophysical signals.

We also note that the high $Z^3$-small $R^3$ tail of the candidate distribution in this search is much less prominent compared to our past Bayesian follow-ups of broad surveys \citep{Martins:2025jnq,McGloughlin:2025eso}.
We explain these differences with the significantly longer coherence time of Stages 0-2 as compared to \cite{McGloughlin:2025eso}.
The longer coherence time reduces the effect short noise-transients can have on our results.

\begin{deluxetable*}{lcccccccccccccccc}
	\tablecaption{Overview of the full hierarchy of searches. For each stage we show the values of the following parameters: the coherent time baseline $T_\mathrm{coh}$ of each segment and the number of segments $N_\mathrm{seg}$; the grid spacings $\delta f,\delta \dot{f}$ and $m_{\text{sky}}$; the average mismatch $\left<\mu \right>$; the parameter space volume searched around each candidate, $\Delta f, \Delta \dot{f}$ and ${\textrm{r}_\textrm{sky}}$. The radius  ${\textrm{r}_\textrm{sky}}$ is expressed in units of the side of the hexagon sky-grid tile of the Stage 0 search (Eq.~\ref{eq:skyGridSpacing});
		the number of candidates searched (N$_{\textrm{in}}$ ) and how many of those survive and make it to the next stage (N$_{\textrm{out}}$). For the Bayesian search stages, we give the number of live points $n_\mathrm{live}$ and proposal strategies (diff and vol. refer to the natively supported proposal strategies of \texttt{Bilby}), and the average proper search volume per candidate in $\langle N_\star \rangle$ \citep{Ashton:2018ure}.
	\label{tab:FUtable}
	}
	\tablehead{
		\colhead{Search} & \colhead{$T_\mathrm{coh}$} & \colhead{$N_\mathrm{seg}$} & \colhead{$\delta f$} & \colhead{$\delta \dot{f}$} & \colhead{$m_{\text{sky}}$} & \colhead{$\left<\mu \right>$ } & \colhead{$\Delta f$} & \colhead{$\Delta \dot{f}$} & \colhead{$ {\textrm{r}_\textrm{sky}\over {d(\mskyFUZero )}}$} &  \colhead{ N$_{\textrm{in}}$ }& \colhead{N$_{\textrm{out}}$} \\
		& hr &  & $\mu{\textrm{Hz}}$ &  {{ $10^{\scriptstyle{-14}}$} Hz/s} & & $10^{-2}$ & $\mu{\textrm{Hz}}$ & {{ $10^{\scriptstyle{-14}}$} Hz/s} &  & &
	}
	\startdata
	Stage 0          & $\TcohFUZero$              & $\NSegFUZero$                & $\dfreqmuHzFUZero$           & $\dfdotfgTenfHzFUZero$             & $\mskyFUZero$              & \meanMismatchFUZero            & \tiny{full range}            & \tiny{full range}                           & \tiny{all-sky}                                                & $\paramtotaltemplates$        & $\NCandFUOne$                \\
	Stage 1          & $\TcohFUOne$               & $\NSegFUOne$                 & $\dfreqmuHzFUOne$            & $\dfdotfgTenfHzFUOne$               & $\mskyFUOne$               & \meanMismatchFUOne             & $\SRfreqmuHzFUOne$           & $\SRfdotFUOne$                              & $\SRskyFUOne$                                                 & $\NCandFUOne$                 & $\NCandFUTwo  $              \\
	\hline
	\hline
	\multicolumn{2}{l}{Bayesian}                                           &                             & \colhead{$n_\mathrm{live}$} & \multicolumn{2}{c}{Proposals }                                            									& $\left<N_\star\right>$ 				&                                &                                             &     & &      \\
	\cline{1-1}\cline{4-7}
	Stage 2          & $\TcohFUTwo$               & $\NSegFUTwo$      & 1000 			& \multicolumn{2}{c}{$\mathrm{DE}_\mathrm{SL}, \mathrm{diff}, \mathrm{vol.}$}                                                                                    &    \NStarFUTwo                         & $\SRfreqmuHzFUTwo$           & $\SRfdotFUTwo$                              & $\SRskyFUTwo$                                                 & $\NCandFUTwo$                 & $\NCandFUThree$              \\
	Stage 3          & \TcohFUThree               & $\NSegFUThree$    &750           	& \multicolumn{2}{c}{diff}                            														 & -\NStarFUThree                         & -                            & -                                           &- 															& $\NCandFUThree$               & $\NCandFUFour $              \\
	\enddata
\end{deluxetable*}

\section{Results}\label{sec:Results}
\subsection{Hardware Injection Recovery}
\label{sec:hardware_injections}

The gravitational wave data contains fake continuous wave signals that are added at the detector level, the hardware injections \citep{Biwer:2016oyg}.
They are used to validate search pipelines and provide a standard benchmark to compare the performance of different searches and methods.

Three hardware injections representing continuous wave signals from isolated neutron stars fall in our search range, specifically those with IDs 3, 5 and 11 \citep{O3_injection_params}.
We recover all three.
Table~\ref{tab:HIRecoveryMonteCarlo} lists the signal parameters and their distances to the associated maximum likelihood estimators found by the Bayesian follow-up at Stage 3, which are all consistent with expectations given the O3a observation span.

The hardware injections with IDs 3 and 5 have been found by other broad searches covering similar parameter ranges as this search, \ie \citet{Steltner:2023cfk,KAGRA:2022dwb,Dergachev:2025hwp,Dergachev:2025ead}, while no other broad survey conducted so far was able to detect the injection with $\mathrm{ID} 11$. This showcases the sensitivity of the search presented here. 

\begin{deluxetable*}{lccccccc}
	\tablecaption{Frequency $f$, frequency derivative $\dot{f}$ and sky position $\alpha, \delta$ of the hardware injections and the offsets between these and the maximum likelihood estimators recovered by our coherent Monte Carlo search using O3a data (Stage $3$). The frequency $f$ is given at reference time \hwinjreftimestageseven~(GPS time).  \label{tab:HIRecoveryMonteCarlo}}
	\tablehead{
		\colhead{ID$_{\rm inj}$} & \colhead{ $f$}  & \colhead{$\dot{f}$} & \colhead{$\alpha$} & \colhead{$\delta$} & \colhead{$\Delta f$} & \colhead{$\Delta \dot{f}$} & \colhead{Sky distance} \\
		& [Hz]            & [Hz/s]              & [hr:m:s]           & [deg:m:s]          & [Hz]                 & [Hz/s]                     & [deg:m:s]
	}
	\startdata
	\hwinjthreeid        & \hwinjthreefreq & \hwinjthreefonedot     & \hwinjthreealpha   & \hwinjthreedelta   & \hwinjthreedfreq     & \hwinjthreedfonedot           & \hwinjthreedsky        \\
	\hwinjfiveid          & \hwinjfivefreq   & \hwinjfivefonedot       & \hwinjfivealpha     & \hwinjfivedelta     & \hwinjfivedfreq       & \hwinjfivedfonedot             & \hwinjfivedsky          \\
	\hwinjelevenid        & \hwinjelevenfreq & \hwinjelevenfonedot     & \hwinjelevenalpha   & \hwinjelevendelta   & \hwinjelevendfreq     & \hwinjelevendfonedot           & \hwinjelevendsky        \\
	\enddata
\end{deluxetable*}

\subsection{The remaining candidate}\label{sec:remaining_candidate}
\begin{figure}
	\includegraphics[width=\columnwidth]{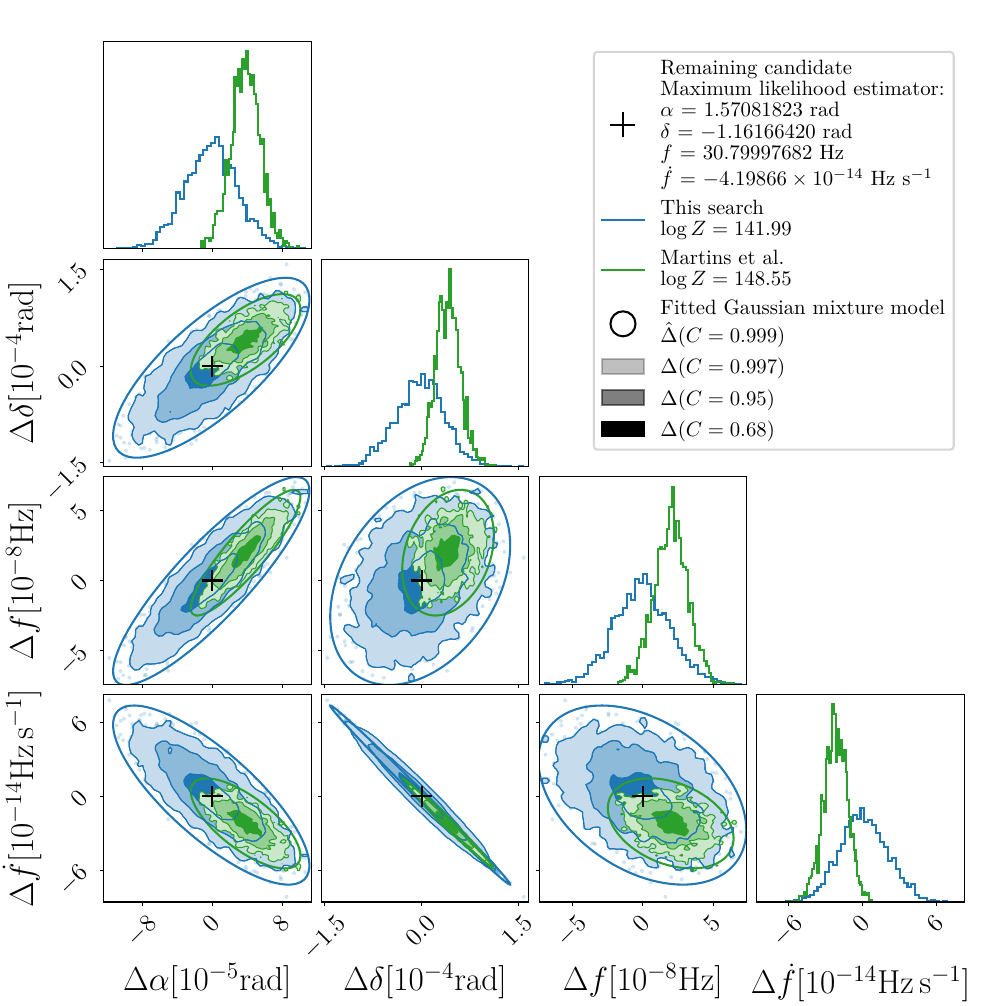}
	\caption{The posterior distribution of the remaining candidate after Stage 3 (blue). We also show the posterior distribution of the remaining candidate found in \citep{Martins:2025jnq} (green). The posteriors coincide. The tighter shape of the posterior in \citet{Martins:2025jnq} is caused by biases in the parameter estimation of that follow-up.}
	\label{fig:remainingcandidateposterior}
\end{figure}

One candidate compatible with the reference population of test signals at Stage 3 remains.
The posterior distribution of the candidate it is shown in Fig.~\ref{fig:remainingcandidateposterior}.
As can be seen, the posterior matches the single remaining candidate that was found and rejected as a defensible signal candidate in the Bayesian follow-up of the ``Deep Einstein@Home Search'' \citep{Steltner:2023cfk} presented in \citet{Martins:2025jnq}.

As discussed in more detail in \citet{Martins:2025jnq}, the candidate arises from known line-like features present in the LLO and LHO data.
The cause of the line in LLO was identified as non-astrophysical only after the data was prepared, and thus was not removed by line-cleaning (see \citet{Steltner:2021qjy}). 
Had it been cleaned, we expect that this candidate would have been rejected by the $\hat{\beta}_{S/GLtL}$-statistic.This candidate illustrates the importance of noise mitigation techniques and rigorous investigations into the causes of noise in the detectors (see also \citet{Covas:2018oik,Davis:2018yrz,LIGO:2021ppb,Capote:2024rmo})

\subsection{Upper Limits}
\label{sec:upperlimits}

\begin{figure*}
	\includegraphics[width=\textwidth]{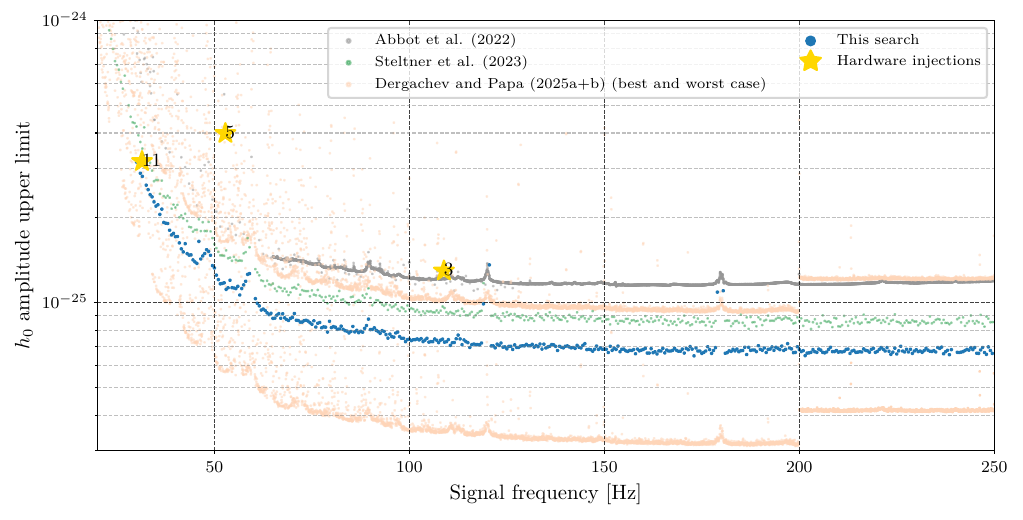}
	\caption{Smallest gravitational wave amplitude $h_0$ that we can exclude from our target population. We compare to other all-sky searches in LIGO O3 data \citep{KAGRA:2022dwb, Steltner:2023cfk,Dergachev:2025hwp,Dergachev:2025ead}. For \citet{KAGRA:2022dwb} only the most stringent upper limits of the different search pipelines used are shown. There are two curves for \citet{Dergachev:2025hwp} and \citet{Dergachev:2025ead} showing the best and worst case (circular and linear polarization) upper limits. The golden stars are the gravitational wave amplitudes $h_0$ of the hardware injections. We recover all three hardware injections within the parameter space range of this search. This search marks the first successful recovery of hardware injection 11 which is consistent with our upper limits set.}
	\label{fig:h0ULs}
\end{figure*}

Following \citet{Steltner:2023cfk} we set upper limits on the gravitational wave amplitude $h_0$ in half-Hz bands. 

There are two exceptions where upper limits are not placed: 1) Known environmental and instrumental disturbances are line-cleaned, i.e. disturbed frequencies are replaced by Gaussian noise in a trade-off to mitigate spectral contamination \citep{Steltner:2021qjy}. This removes (parts of the) signals. Half-Hz bands in which 90\% recovery can not be achieved because of this are excluded from our upper limits. 2) Upper limits are also not set in \fiftyMHzband{} bands which are classified as disturbed in the visual inspection process, because in these bands the statistical properties of the search results are hard to model.

The upper limits are shown in Fig.\ref{fig:h0ULs}. The upper limits, as well as the half-Hz and \fiftyMHzband{} bands in which upper limits are not placed, are given in text form in the supplemental material \citep{O3ASBu-AEI}.

\begin{figure}
	\centering
	\includegraphics[width=\columnwidth]{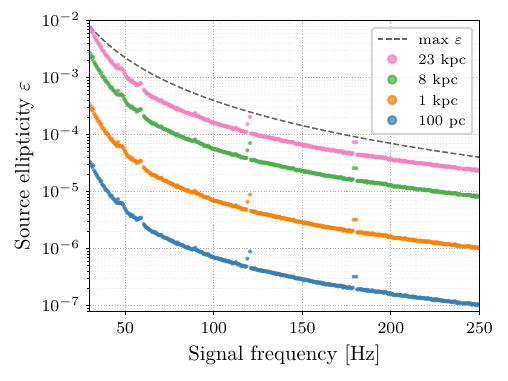}
	\caption{Upper limits on the neutron star ellipticity $\varepsilon$ at different distances. The dashed line shows the maximum ellipticity probed due to the maximum spin-down of this search. The 23 kpc marks the maximum distance from Earth within which the majority of galactic neutron stars are expected to lie.}
	\label{fig:epsilonULs}
\end{figure}

The upper limits on $h_0$ can be translated in upper limits on the ellipticity $\varepsilon$ of a source modelled as a triaxial ellipsoid spinning around a principal moment of inertia axis ${I}$ at a distance $d$:
\begin{equation}
	\begin{split}
		\varepsilon^{\textrm{UL}} & =1.4 \times 10^{-6} ~\left( {h_0^{\textrm{UL}}\over{1.4\times10^{-25}}}\right ) \times                                                    \\
		                          & \left ( {d\over{1~\textrm{kpc}}}\right ) \left ({{\textrm{170~Hz}}\over f} \right )^2 \left ({10^{38}~{\textrm{kg m}}^2\over I} \right ). \\
	\end{split}
	\label{eq:epsilon}
\end{equation}
Figure \ref{fig:epsilonULs} shows the upper limits on the ellipticity $\varepsilon$ for different distances.

The value of neutron-star ellipticities is very uncertain, in fact even the maximum deformation that a neutron star crust can support is unknown, with models predicting maximum ellipticities on the order of $10^{-5}$ \citep{JohnsonMcDaniel:2012wg, Morales:2022wxs} down to at most $10^{-9}$ \citep{Bhattacharyya:2020paf,Gittins:2021zpv}. This search probes ellipticities of  $10^{-7}$ level for neutron stars spinning about $8$ ms within $100$ pc, that comfortably sits in this range.

\begin{figure}
	\includegraphics[width=\columnwidth]{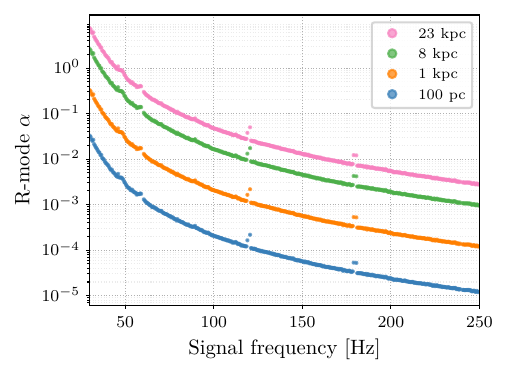}
	\caption{Upper limits on the r-mode amplitude $\alpha$ at different distances.}
	\label{fig:rmodeULs}
\end{figure}

Another possible emission mechanism are r-modes, unstable toroidal fluid oscillations driven by the Coriolis force, emitting at $\approx 4/3$ of the spin-frequency \citep{Andersson:1997xt,Friedman:1997uh,Lindblom:1998wf}. We translate the upper limits on the gravitational wave amplitude $h_0$ in upper limits on the r-mode amplitude \citep{Owen:2010ng}:

\begin{equation}
	\alpha^{\textrm{UL}} = 0.028
	\left( \frac{h_0^{\textrm{UL}}}{\num{1e-24}} \right)
	\left( \frac{d}{1 \mathrm{kpc}} \right)
	\left( \frac{\SI{100}{\hertz}}{f} \right)^3.
\end{equation}

Figure \ref{fig:rmodeULs} shows the upper limits on the r-mode amplitude $\alpha$ for different distances $d$. Young neutron stars are promising emitters of r-modes, and the values of r-mode amplitude probed here are not implausible for young neutron stars \citep{Bondarescu:2008qx}.

\section{Conclusions}\label{sec:Conclusions}

We present the results of a search for continuous gravitational wave emission from all over the sky, with frequencies $\SI{\paramfmin}{\hertz} \leq f \leq \SI{\paramfmax}{\hertz}$ \text{and spin-downs} $-2.7 \times 10^{-9}\, \text{Hz/s} \leq \dot{f} \leq 0.2 \times 10^{-9}$ Hz/s using LIGO O3 public data. This search improves in sensitivity with respect to the current most sensitive broad surveys \citep{Steltner:2023cfk,McGloughlin:2025eso} on similar parameter spaces, by $\approx 25\%$.

This is the most sensitive all-sky search in the search region considered. All hardware injections within the search range are recovered. In particular, this search marks the first successful recovery of hardware injection 11.

No plausible astrophysical candidate survives the large-scale follow-up procedure. Upper limits are set on the gravitational wave amplitude $h_0$. The most stringent upper limits are at \mostStringentULFreq{} Hz with $h_{0} = \mostStringentUL$, at the 90\% confidence level, which translates to excluding neutron stars with a spin period $P < 10\, \rm ms$ with ellipticities $\epsilon >1.5 \times 10^{-7}(d/100\,\rm pc)$. At $100$ pc our results begin to exclude ellipticities less than $10^{-6}$ for neutron starts with spin periods below 22 ms. 

We look forward to probing even smaller deformations with the upcoming public release of the LIGO O4 data. 

\section*{Acknowledgments}
\begin{acknowledgments}
	We thank the Einstein@Home volunteers, without the support of whom this search could not have happened. We acknowledge the use of the HPC systems Raven and Viper at the Max Planck Computing and Data Facility. We acknowledge the use of Topcat \citep{2005ASPC..347...29T} and scikit-learn \citep{pedregosa2011scikit}. This research has used data or software obtained from the Gravitational Wave Open Science Center (gwosc.org), a service of LIGO Laboratory, the LIGO Scientific Collaboration, the Virgo Collaboration and KAGRA.
\end{acknowledgments}

\bibliography{references}

\begin{thebibliography}{}
\expandafter\ifx\csname natexlab\endcsname\relax\def\natexlab#1{#1}\fi
\providecommand{\url}[1]{\href{#1}{#1}}
\providecommand{\dodoi}[1]{doi:~\href{http://doi.org/#1}{\nolinkurl{#1}}}
\providecommand{\doeprint}[1]{\href{http://ascl.net/#1}{\nolinkurl{http://ascl.net/#1}}}
\providecommand{\doarXiv}[1]{\href{https://arxiv.org/abs/#1}{\nolinkurl{https://arxiv.org/abs/#1}}}

\bibitem[{Abbott {et~al.}(2017)}]{LIGOScientific:2017wva}
Abbott, B.~P., {et~al.} 2017, Phys. Rev. D, 96, 122004,
  \dodoi{10.1103/PhysRevD.96.122004}

\bibitem[{Abbott {et~al.}(2022)}]{KAGRA:2022dwb}
Abbott, R., {et~al.} 2022, Phys. Rev. D, 106, 102008,
  \dodoi{10.1103/PhysRevD.106.102008}

\bibitem[{{Anderson}(2004)}]{Boinc2}
{Anderson}, D.~P. 2004, in Proceedings of the Fifth IEEE/ACM International
  Workshop on Grid Computing (GRID04), 4--10

\bibitem[{{Anderson} {et~al.}(2006){Anderson}, {Christensen}, \&
  {Allen}}]{Boinc3}
{Anderson}, D.~P., {Christensen}, C., \& {Allen}, B. 2006, in Proceedings of
  the 2006 ACM/IEEE conference on Supercomputing, 126--136

\bibitem[{Andersson(1998)}]{Andersson:1997xt}
Andersson, N. 1998, Astrophys. J., 502, 708, \dodoi{10.1086/305919}

\bibitem[{Ashton \& Prix(2018)}]{Ashton:2018ure}
Ashton, G., \& Prix, R. 2018, Phys. Rev. D, 97, 103020,
  \dodoi{10.1103/PhysRevD.97.103020}

\bibitem[{Ashton {et~al.}(2019)}]{Ashton:2018jfp}
Ashton, G., {et~al.} 2019, The Astrophysical Journal Supplement Series, 241,
  27, \dodoi{10.3847/1538-4365/ab06fc}

\bibitem[{Behnke {et~al.}(2015)Behnke, Papa, \& Prix}]{Behnke:2014tma}
Behnke, B., Papa, M.~A., \& Prix, R. 2015, Phys. Rev. D, 91, 064007,
  \dodoi{10.1103/PhysRevD.91.064007}

\bibitem[{Bhattacharyya(2020)}]{Bhattacharyya:2020paf}
Bhattacharyya, S. 2020, {The permanent ellipticity of the neutron star in PSR
  J1023+0038}, \dodoi{10.1093/mnras/staa2304}

\bibitem[{Biwer {et~al.}(2017)}]{Biwer:2016oyg}
Biwer, C., {et~al.} 2017, Physical Review D, 95, 062002,
  \dodoi{10.1103/PhysRevD.95.062002}

\bibitem[{Bondarescu {et~al.}(2009)Bondarescu, Teukolsky, \&
  Wasserman}]{Bondarescu:2008qx}
Bondarescu, R., Teukolsky, S.~A., \& Wasserman, I. 2009, Phys. Rev. D, 79,
  104003, \dodoi{10.1103/PhysRevD.79.104003}

\bibitem[{Buchner(2022)}]{Buchner:2022}
Buchner, J. 2022, Physical Sciences Forum, 5, \dodoi{10.3390/psf2022005046}

\bibitem[{Buikema {et~al.}(2020)}]{aLIGO:2020wna}
Buikema, A., {et~al.} 2020, Phys. Rev. D, 102, 062003,
  \dodoi{10.1103/PhysRevD.102.062003}

\bibitem[{Capote {et~al.}(2025)}]{Capote:2024rmo}
Capote, E., {et~al.} 2025, Phys. Rev. D, 111, 062002,
  \dodoi{10.1103/PhysRevD.111.062002}

\bibitem[{Covas {et~al.}(2018)}]{Covas:2018oik}
Covas, P., {et~al.} 2018, Phys. Rev. D, 97, 082002,
  \dodoi{10.1103/PhysRevD.97.082002}

\bibitem[{Cutler \& Schutz(2005)}]{Cutler:2005hc}
Cutler, C., \& Schutz, B.~F. 2005, Phys. Rev. D, 72, 063006,
  \dodoi{10.1103/PhysRevD.72.063006}

\bibitem[{Davis {et~al.}(2019)Davis, Massinger, Lundgren, Driggers, Urban, \&
  Nuttall}]{Davis:2018yrz}
Davis, D., Massinger, T.~J., Lundgren, A.~P., {et~al.} 2019, Class. Quant.
  Grav., 36, 055011, \dodoi{10.1088/1361-6382/ab01c5}

\bibitem[{Davis {et~al.}(2021)}]{LIGO:2021ppb}
Davis, D., {et~al.} 2021, Class. Quant. Grav., 38, 135014,
  \dodoi{10.1088/1361-6382/abfd85}

\bibitem[{Dergachev \& Papa(2025{\natexlab{a}})}]{Dergachev:2025hwp}
Dergachev, V., \& Papa, M.~A. 2025{\natexlab{a}}, Phys. Rev. D, 112, 042005,
  \dodoi{10.1103/psrl-y44w}

\bibitem[{Dergachev \& Papa(2025{\natexlab{b}})}]{Dergachev:2025ead}
---. 2025{\natexlab{b}}.
\newblock \doarXiv{2507.12161}

\bibitem[{Friedman \& Morsink(1998)}]{Friedman:1997uh}
Friedman, J.~L., \& Morsink, S.~M. 1998, Astrophys. J., 502, 714,
  \dodoi{10.1086/305920}

\bibitem[{Gittins \& Andersson(2021)}]{Gittins:2021zpv}
Gittins, F., \& Andersson, N. 2021, Mon. Not. Roy. Astron. Soc., 507, 116,
  \dodoi{10.1093/mnras/stab2048}

\bibitem[{Jaranowski {et~al.}(1998)Jaranowski, Krolak, \&
  Schutz}]{Jaranowski:1998qm}
Jaranowski, P., Krolak, A., \& Schutz, B.~F. 1998, Phys. Rev., D58, 063001,
  \dodoi{10.1103/PhysRevD.58.063001}

\bibitem[{Johnson-McDaniel \& Owen(2013)}]{JohnsonMcDaniel:2012wg}
Johnson-McDaniel, N.~K., \& Owen, B.~J. 2013, Phys. Rev. D, 88, 044004,
  \dodoi{10.1103/PhysRevD.88.044004}

\bibitem[{Keitel(2016)}]{Keitel:2015ova}
Keitel, D. 2016, Phys. Rev. D, 93, 084024, \dodoi{10.1103/PhysRevD.93.084024}

\bibitem[{LIGO \& Virgo(2022)}]{O3_injection_params}
LIGO, S.~C., \& Virgo, P. 2022,
  \url{https://www.gw-openscience.org/O3/O3April1_injection_parameters/}

\bibitem[{Lindblom {et~al.}(1998)Lindblom, Owen, \& Morsink}]{Lindblom:1998wf}
Lindblom, L., Owen, B.~J., \& Morsink, S.~M. 1998, Phys. Rev. Lett., 80, 4843

\bibitem[{Martins {et~al.}(2025)Martins, Papa, Steltner, Prix, \&
  Vidal}]{Martins:2025jnq}
Martins, J., Papa, M.~A., Steltner, B., Prix, R., \& Vidal, P.~C. 2025.
\newblock \doarXiv{2508.18204}

\bibitem[{McGloughlin {et~al.}(2025{\natexlab{a}})McGloughlin, Martins,
  Steltner, Papa, Eggenstein, Machenschalk, Prix, \& Bensch}]{O3ASBu-AEI}
McGloughlin, B., Martins, J., Steltner, B., {et~al.} 2025{\natexlab{a}},
  {Supplemental materials to the paper {\it{Einstein@Home all-sky ``bucket"
  search for continuous gravitational waves in LIGO O3 public data}}},
  \url{www.aei.mpg.de/continuouswaves/EaH-O3ASBu}

\bibitem[{McGloughlin {et~al.}(2025{\natexlab{b}})McGloughlin, Steltner,
  Martins, Papa, Eggenstein, Ming, Machenschalk, Prix, \&
  Bensch}]{McGloughlin:2025eso}
McGloughlin, B., Steltner, B., Martins, J., {et~al.} 2025{\natexlab{b}}.
\newblock \doarXiv{2508.20073}

\bibitem[{Morales \& Horowitz(2022)}]{Morales:2022wxs}
Morales, J.~A., \& Horowitz, C.~J. 2022, {Neutron Star Crust Can Support A
  Large Ellipticity}, \dodoi{{10.1093/mnras/stac3058}}

\bibitem[{Owen(2010)}]{Owen:2010ng}
Owen, B.~J. 2010, Phys. Rev. D, 82, 104002, \dodoi{10.1103/PhysRevD.82.104002}

\bibitem[{Owen {et~al.}(1998)Owen, Lindblom, Cutler, Schutz, Vecchio, \&
  Andersson}]{Owen:1998xg}
Owen, B.~J., Lindblom, L., Cutler, C., {et~al.} 1998, Phys. Rev. D, 58, 084020,
  \dodoi{10.1103/PhysRevD.58.084020}

\bibitem[{Pedregosa {et~al.}(2011)Pedregosa, Varoquaux, Gramfort, Michel,
  Thirion, Grisel, Blondel, Prettenhofer, Weiss, Dubourg,
  {et~al.}}]{pedregosa2011scikit}
Pedregosa, F., Varoquaux, G., Gramfort, A., {et~al.} 2011, Journal of machine
  learning research, 12, 2825

\bibitem[{Pletsch(2010)}]{Pletsch:2010xb}
Pletsch, H.~J. 2010, Phys. Rev. D, 82, 042002,
  \dodoi{10.1103/PhysRevD.82.042002}

\bibitem[{Prix \& Krishnan(2009)}]{Prix:2009tq}
Prix, R., \& Krishnan, B. 2009, Class. Quant. Grav., 26, 204013,
  \dodoi{10.1088/0264-9381/26/20/204013}

\bibitem[{Skilling(2004)}]{Skilling:2004pqw}
Skilling, J. 2004, AIP Conference Proceedings, 735, 395,
  \dodoi{10.1063/1.1835238}

\bibitem[{Speagle(2020)}]{Speagle:2019ivv}
Speagle, J.~S. 2020, Monthly Notices of the Royal Astronomical Society, 493,
  3132, \dodoi{10.1093/mnras/staa278}

\bibitem[{Steltner {et~al.}(2022{\natexlab{a}})Steltner, Menne, Papa, \&
  Eggenstein}]{Steltner:2022aze}
Steltner, B., Menne, T., Papa, M.~A., \& Eggenstein, H.-B. 2022{\natexlab{a}},
  {Phys. Rev. D}, {106}, {104063}, \dodoi{10.1103/PhysRevD.106.104063}

\bibitem[{Steltner {et~al.}(2022{\natexlab{b}})Steltner, Papa, \&
  Eggenstein}]{Steltner:2021qjy}
Steltner, B., Papa, M.~A., \& Eggenstein, H.-B. 2022{\natexlab{b}}, Phys. Rev.
  D, 105, 022005, \dodoi{10.1103/PhysRevD.105.022005}

\bibitem[{Steltner {et~al.}(2023)Steltner, Papa, Eggenstein, Prix, Bensch,
  Allen, \& Machenschalk}]{Steltner:2023cfk}
Steltner, B., Papa, M.~A., Eggenstein, H.~B., {et~al.} 2023, Astrophys. J.,
  952, 55, \dodoi{10.3847/1538-4357/acdad4}

\bibitem[{Steltner {et~al.}(2021)}]{Steltner:2020hfd}
Steltner, B., {et~al.} 2021, The Astrophysical Journal, 909, 79,
  \dodoi{10.3847/1538-4357/abc7c9}

\bibitem[{{Taylor}(2005)}]{2005ASPC..347...29T}
{Taylor}, M.~B. 2005, in Astronomical Society of the Pacific Conference Series,
  Vol. 347, Astronomical Data Analysis Software and Systems XIV, ed.
  P.~{Shopbell}, M.~{Britton}, \& R.~{Ebert}, 29

\bibitem[{Zhu {et~al.}(2020)Zhu, Baryakhtar, Papa, Tsuna, Kawanaka, \&
  Eggenstein}]{Zhu:2020tht}
Zhu, S.~J., Baryakhtar, M., Papa, M.~A., {et~al.} 2020, Phys. Rev. D, 102,
  063020, \dodoi{10.1103/PhysRevD.102.063020}

\end{thebibliography}

\end{document}